\documentclass[11pt]{article}

\usepackage{a4wide}
\usepackage{mathrsfs}
\usepackage[T1]{fontenc}
\usepackage{mathpazo}
\usepackage{setspace}
\usepackage{amsfonts}
\usepackage{amssymb}
\usepackage{amsmath}
\usepackage{epsfig}
\usepackage{latexsym}
\usepackage{color}
\usepackage{nicefrac}
\usepackage[latin1]{inputenc}

\author{
  \begin{minipage}{.97\linewidth}
    \vspace{1cm}
    \begin{center}
      \begin{small}
        \textbf{P.M. Petropoulos}\footnote{marios@cpht.polytechnique.fr}
      \end{small}
    \end{center}
    \vspace{0.5cm}
    \hspace{2cm}\begin{minipage}{.7\linewidth}
     {\it \begin{footnotesize}
     \begin{center}
         Centre de Physique Th\'eorique\\
        Ecole Polytechnique,  CNRS UMR 7644\\
        91128 Palaiseau Cedex, France
        \end{center}
     \end{footnotesize}}
    \end{minipage}
    \vspace{0.5cm}
  \end{minipage}
}

\date{\today}

\title{\vspace{0.75cm}
 \boldmath \begin{huge}
    \textbf{Geometric flows and applications}
  \end{huge} \unboldmath
}

\begin{document}

\begin{titlepage}
  \maketitle
  \thispagestyle{empty}

  \vspace{-9.5cm}
  \begin{flushright}
    CPHT-PC012.0210
  \end{flushright}

  \vspace{10cm}

  \begin{center}
    \textsc{Abstract}\\
  \end{center}
I present some applications of geometric flows in string theory and gravity. In some circumstances  time evolution in string theory can be approximately identified with Ricci-flow parametric evolution of spatial sections. In four dimensions, homogeneous, self-dual, gravitational instantons of general relativity evolve in time exactly as geometric flows of homogeneous three-manifolds. For non-relativistic versions of gravity, this property persists in any dimension, under the assumption of detailed-balance condition. 

\vspace{3cm} \noindent To appear in the proceedings of the
\textsl{Corfu09 Summer Institute on Elementary Particle Physics}.

\end{titlepage}

\onehalfspace

\tableofcontents


\section{Introduction}

Ricci flows were introduced by R. Hamilton in 1982 as the pillar of a programme designed to prove Poincar\'e's (1904) and Thurston's (mid-seventies) conjectures. This programme culminated in 2002--2003 with the actual proof by Perel'man \cite{perel:0203}. Meanwhile, many applications have been found in physics, particularly in subjects related to gravity. The aim of the present notes is to overview some of these applications. I will discuss the issue of time dependence in string theory, as well as an intriguing relation between geometric flows of three-dimensional homogeneous spaces and four-dimensional self-dual gravitational instantons of general relativity. Finally, I will make a little detour in non-relativistic theories of gravity, where the filiation of geometric flows with gravitational instantons holds in any dimension, under certain conditions that will be made precise in due time. By lack of space, this exhibition is neither complete nor exhaustive. It leaves aside several interesting subjects such as tachyon condensation, black-hole physics or D-branes, where Ricci flows have occasionally appeared. 

\section{A glance on geometric flows}

\subsection{Generalities}

Geometric flows describe an irreversible parametric evolution of the metric on a Riemannian manifold, under a general non-linear parabolic equation, given a symmetric tensor $S_{ij}[g]$:
\begin{equation}
\nonumber
\frac{\partial g_{ij}}{\partial t}=S_{ij}.
\end{equation}
A plethora of flows have been investigated 
based on various tensors
and leading to diverse behaviours. 
The Ricci flow corresponds to the choice $S_{ij}=-R_{ij}$. For this flow, the volume is not preserved:
   \begin{equation}\label{voldep}
\frac{\mathrm{d}V}{\mathrm{d}t}
=\frac{1}{2} \int \mathrm{d}^Dx \sqrt{\det g}    g^{ij}
\frac{\partial g_{ij}}{\partial t} =-\frac{1}{2} \int \mathrm{d}^Dx \sqrt{\det g} R.
 \end{equation}
As a consequence, positive curvature induces contraction, whereas negative curvature leads to expansion. A noticeable feature of the Ricci flow is that it preserves Killing vectors: the isometry group  can only grow or remain unaltered. Many other interesting properties and relations to its original goal can be found in \cite{chow:2004}. 

\subsection{Three-dimensional homogeneous spaces}\label{homspa}

Ricci flows turned particularly useful in three dimensions, where Poincar\'e and Thurston's conjectures were originally  formulated. The homogeneous three-manifolds played a distinguished r\^ole in that respect, and were extensively analysed, mostly with asymptotic methods (see e.g. \cite{Isenberg:1992}). 

Homogeneous manifolds admit a group $G$ acting transitively. In three dimensions, the classification of these spaces is straightforward (see e.g. \cite{Scott:1983}), especially when the action of $G$ is simply transitive. In this case $\dim G=3$, the group is of  Bianchi type, and there are three linearly independent Killing vectors, tangent to $\mathcal{M}_3$, satisfying  
\begin{equation}
\nonumber
\left[\xi_i, \xi_j\right] =c^i_{\hphantom{i}jk} \xi_k, 
\end{equation}
where $c^i_{\hphantom{i}jk} $ are the structure constants\footnote{There are nine Bianchi groups split into two classes, depending on whether $c^i_{\hphantom{i}ij}$ vanishes (class A, \emph{unimodular}) or not (class B, \emph{non-unimodular}). 
}. The latter also appear in the Cartan structure equations for the left-invariant Cartan--Maurer forms $ \sigma^i$:
\begin{equation}   
\nonumber
\mathrm{d}\sigma^i = \frac{1}{2} c^i_{\hphantom{i}jk}\sigma^j \wedge \sigma^k .  
\end{equation}   
The most general invariant metric on  $\mathcal{M}_3$ is of the form
\begin{equation}   
\label{met3}
\mathrm{d}s^2= \gamma_{ij}   \sigma^i \sigma^j,
\end{equation}   
where $\gamma_{ij} $ are coordinate-independent. For unimodular Bianchi groups, it can always be chosen diagonal.

\subsection{Ricci flow on the three-sphere}\label{s3}

Let us concentrate for concreteness on the Bianchi IX class, where $G\equiv SU(2)$ and $\mathcal{M}_3$ is a three-sphere. The structure constants are thus $ c^i_{\hphantom{i}jk} =\epsilon^i_{\hphantom{i}jk}$, and the diagonal ansatz for the metric,  $\gamma_{ij} =\gamma_i  \delta_{ij} $,
captures the most general situation. This metric is $SU(2)$-invariant, although the symmetry can be enhanced to $SU(2)\times U(1)$, when two $\gamma$s are equal, and to $SU(2)\times SU(2)$, when the  
$\gamma$s are all equal. This latter case corresponds to the round sphere. 

As was noticed in \cite{Sfetsos:2006}, and further exploited in \cite{Bakas:2006bz}, the Ricci-flow equations for this metric, when the $\gamma_i$s depend on a flow parameter $t$, are the Darboux--Halphen  system studied in the nineteenth century:
\begin{equation}\label{halp}
\begin{cases}
    \dot{\Omega}^1 = \Omega^2 \Omega^3 - \Omega^1 \left(\Omega^2
      + \Omega^3 \right),   \\
    \dot{\Omega}^2 = \Omega^3 \Omega^1 - \Omega^2 \left(\Omega^3
      + \Omega^1 \right),   \\
    \dot{\Omega}^3 = \Omega^1 \Omega^2 - \Omega^3 \left(\Omega^1 +
      \Omega^2 \right).
  \end{cases}
\end{equation}
Here I have traded $\gamma_i$ for $\Omega^i=\gamma_j\gamma_k$, and the dot stands for 
 $\nicefrac{\mathrm{d}}{\mathrm{d}T}=\sqrt{\Omega^1\Omega^2 \Omega^3}\,\nicefrac{\mathrm{d}}{\mathrm{d}t}$. In the fully anisotropic case, this system is integrable using quasi-modular forms \cite{Darboux, halph}. As shown in \cite{Bakas:2006bz}, for finite positive initial conditions, the flow remains positive with universal late-time behaviour $\Omega^{i}\approx \nicefrac{1}{T}$. The $S^3$ flows toward a round sphere with vanishing radius\footnote{This is not surprising, according to the general property quoted previously (Eq. (\ref{voldep})). It can nevertheless be avoided, either with a normalized flow, or with a torsionfull connection.}. Other Bianchi classes exhibit different behaviours (absence of convergence, pancake degeneracies, cigar degeneracies~\dots) as discussed in Ref. \cite{Isenberg:1992}.

\section{Sigma-models, renormalization-group flows and cosmology}

String propagation in arbitrary backgrounds is described in terms of a non-linear sigma-model. Generally, the latter is not scale-invariant, and the background fields flow according to the renormalization-group  two-dimensional mass scale $\mu$: 
  \begin{equation}
  \nonumber
  \frac{\partial }{\partial \ln \mu^{-1}}g_{MN}=-\beta\left[g_{MN}\right],
\end{equation}
and similarly for other background fields, where the beta-functions are determined order by order in $\alpha'$. When the background is metric only, the worldsheet action reads:
  \begin{equation}
  \nonumber
  S=\frac{1}{2\pi\alpha'}\int \mathrm{d}^2z \, g_{MN}(x) \partial x^M\bar\partial x^N,
\end{equation}
and consequently \cite{Friedan:1980jm}
\begin{equation}
\nonumber
\beta\left[g_{MN}\right]\equiv R_{MN} + \mathrm{O}(\alpha').
\end{equation}
Introducing the renormalization-group time $t=-\ln \mu$, the renormalization-group evolution is a Ricci flow with past corresponding to the ultra-violet and future to the infrared. Fixed points of the Ricci flow correspond to two-dimensional conformal field theories.

The original aim of string cosmology is to settle time-dependent string backgrounds, i.e. solve the equations $\beta=0$ with background fields like $g_{MN}$ depending on the sigma-model time-like coordinate $x^0$. This is not an easy task, in general, because (\romannumeral1) string theory does not allow for arbitrary matter, (\romannumeral2) around the Big-Bang, high curvature requires exact (in $\alpha'$) solutions, and (\romannumeral3) after the inflation era, the dilaton potential calls for higher-$g_{\mathrm{s}}$ corrections.  

An alternative approach to critical string theory on genuine spacetime backgrounds is the renormalization-group-flowing string on purely space-like environments. Indeed, off-criticality generates a Liouville mode that may eventually play the r\^ole of time coordinate in the sigma-model \cite{Das:1990, Sen:1990}. This suggests that renormalization-group evolution of a purely space-like background may resemble time evolution. Hence, Ricci flow could mimic cosmological evolution with fixed points being steady-state universes. 

These ideas, although appealing, are not a priori justified. Indeed, at the technical level, renormalization-group flow equations are first-order in time: 
 \begin{equation}\label{bij}
\frac{\partial g_{ij}}{\partial t}=-\beta \left[g_{ij}\right],
\end{equation}
whereas genuine evolution is second-order:
 \begin{equation}\label{bmunu}
\beta \left[g_{\mu\nu}\right]=0, 
\end{equation}
and similarly for the other background fields (greek indices are spacetime and latin purely spatial). Thus, the identification of the renormalization-group time with genuine time cannot hold exactly.

The above problem was originally discussed in \cite{Tseytlin:1992ye}. The general analysis was presented in
\cite{Schmidhuber:1994bv}, where it was shown that the dissipative nature of the dilaton helps transforming second-order equations into first-order, as a consequence of friction. I can illustrate this in the example of three-spheres discussed in Sec. \ref{s3}.
Consider a sigma-model where, besides some internal conformal field theory, there is a three-dimensional Bianchi IX target space. The renormalization-group flow for the latter is the Ricci flow discussed in Sec. \ref{s3}.   
In the simplest, isotropic, situation where $\gamma_i(t) =\exp 2\sigma(t) \ \forall i$, Eqs. (\ref{bij}) read: 
 \begin{equation}
 \label{first}
\sigma'=- \frac{\mathrm{e}^{-2\sigma}}{4}, 
\end{equation}
where the prime stands for $\nicefrac{\mathrm{d}}{\mathrm{d}t}$.
The target-space of this sigma-model can be promoted to a genuine spacetime with metric
 \begin{equation}
 \nonumber
\mathrm{d}s^2 =- \mathrm{d}t^2+
  \sum_i\gamma_i \left(\sigma^i\right)^2.
\end{equation}
An extra worldsheet field, $t$, is introduced and,
in order to maintain criticality, a time-dependent dilaton $\Phi(t)$ must also be added. Requiring that beta-functions vanish (Eq. (\ref{bmunu})), leads to second-order equations:
 \begin{equation}\label{C}
 \begin{cases}
Q'=-\frac{3}{2}(\sigma')^2,\\
\sigma''+2Q\sigma'=-\frac{\mathrm{e}^{-2\sigma}}{4},
\end{cases}
\end{equation}
where $Q\equiv-\Phi'+\frac{3}{2}\sigma'$. The field $Q$ plays the r\^ole of a friction force, which dominates in the vicinity of the fixed point\footnote{The fixed point is here a sphere with vanishing radius. The addition of an antisymmetric-tensor background field allows for the radius to remain finite (see e.g. \cite{Bakas:2006bz}).}. In this regime, Eqs. (\ref{first}) and   (\ref{C}) become equivalent, as advertised.  
%

\section{Gravitational instantons and geometric flows}


Gravitational instantons are non-singular, finite-action solutions of Einstein's equations in vacuum or with a cosmological constant. They have been thoroughly studied in the past with, among others, the aim to be used in the description of quantum-gravity transitions (see e.g. \cite{Eguchi:1980jx} for a review and references).  Solving Einstein's equations is a hard problem, unless some specific mini-superspace ansatz is made. It is often assumed, for instance, that (\romannumeral1) the four-dimensional space 
$\mathcal{M}_4 $ is topologically $\mathbb{R} \times\mathcal{ M}_3$, (\romannumeral2) the spatial sections of the foliation $\mathcal{ M}_3$ are homogeneous of Bianchi type $G$, and (\romannumeral3) the curvature two-form is self-dual (anti-self-duality is reached upon parity or time reversal). The latter assumption is inspired from the self-dual Yang--Mills reductions and automatically ensures Ricci flatness. 

With the above assumptions at hand, one can take
\begin{equation}
\label{met}
\mathrm{d}s^2=\mathrm{d}t^2 + g_{ij} \sigma^i  \sigma^j ,
\end{equation}
where the $\sigma^i$s are the Cartan--Maurer invariant forms introduced in Sec. \ref{homspa}.
Adapting the splitting $SU(2)_{\mathrm{sd}}\times SU(2)_{\mathrm{asd}}$ (self-dual and anti-self-dual factors) of  the $SO(4)$ group acting on the tangent space,  
 to the slicing of $\mathcal{M}_4$ into the homogeneous space $\mathcal{M}_3$ and the orthogonal direction $t$, the curvature two-form can be reduced as: 
 \begin{equation}
 \nonumber
\mathcal{S}_i=\frac{1}{2}\left(\mathcal{R}_{0i} + \frac{1}{2} \epsilon_{ijk}\mathcal{R}^{jk}\right), \quad\mathcal{A}_i=\frac{1}{2}\left(\mathcal{R}_{0i} - \frac{1}{2} \epsilon_{ijk}\mathcal{R}^{jk}\right). 
\end{equation}
The self-dual part of the curvature $\mathcal{S}_i$ is a vector of $SU(2)_{\mathrm{sd}}$ and a singlet under 
$ SU(2)_{\mathrm{asd}}$ and vice-versa for the anti-self-dual piece $\mathcal{A}_i$. In terms of the corresponding Levi--Civita connection, reduced as 
\begin{equation}
\nonumber
\Sigma_i=\frac{1}{2}\left(\omega_{0i} + \frac{1}{2} \epsilon_{ijk}\omega^{jk}\right),\quad
A_i=\frac{1}{2}\left(\omega_{0i} - \frac{1}{2} \epsilon_{ijk}\omega^{jk}\right),
\end{equation}
the curvature reads:
\begin{equation}
\nonumber
\mathcal{S}_i= \mathrm{d} \Sigma_i -\epsilon_{ijk} \Sigma^j \wedge  \Sigma^k, \quad
  \mathcal{A}_i= \mathrm{d} A_i +\epsilon_{ijk} A^j \wedge A^k.
\end{equation}

Self-dual gravitational instantons are obtained by imposing $ \mathcal{A}_i=0$, which is solved by any flat connection $A_i$. The classification of self-dual gravitational instantons amounts therefore to finding all possible flat $SU(2)$ connections over $G$. Setting 
\begin{equation}
\label{1stord}
A_i=\frac{\lambda_{ij}}{2}\sigma^j,
\end{equation}
the equation $ \mathcal{A}_i=0$ reads:
\begin{equation}
\label{flat}
\lambda_{i\ell}c^{\ell}_{\hphantom{\ell}jk}+\epsilon_{imn}
\lambda^{m}_{\hphantom{n}[j}\lambda^{n}_{\hphantom{n}k]}=0,
\end{equation}
which has as many inequivalent solutions as algebra homomorphisms $G\to SU(2)$   \cite{Bourliot:2009fr,Bourliot:2009ad}. For each solution $\lambda_{ij}$, Eqs. (\ref{1stord}) provide a different set of first-order equations\footnote{The  $\lambda_{ij}$'s appear in fact as first integrals of the original second-order equations $ \mathcal{A}_i=0$ (see for instance \cite{Gibbons:1979xn} for a more complete overview). The reader might be puzzled at this stage by the fact that, barring global issues, if $ \mathcal{A}_i=0$, one can always find an $SU(2)_{\mathrm{asd}}$ local transformation such that the connection $A_i$ vanishes in the rotated frame. Such a transformation spoils, however, the invariant nature of the original frame. Our viewpoint here is to keep this frame and search for flat anti-self-dual parts in the connection.} that can be solved to deliver distinct classes of instantons.

The above procedure can be successfully applied to all Bianchi classes and allows to obtain  various Ricci-flat gravitational instantons. In the case of Bianchi IX, for instance, the two non-equivalent homomorphisms (the trivial one and the isomorphism) lead respectively to two sets of equations. For a diagonal metric ansatz $g_{ij}=\gamma_i\gamma_j\delta_{ij} $, these sets are respectively the Lagrange 
system ($\Omega^i=\gamma_j\gamma_k$ as previously)
\begin{equation}
\nonumber
\dot{\Omega}^1= -\Omega^2 \Omega^3,\quad
   \dot{\Omega}^2= -\Omega^3 \Omega^1,\quad
   \dot{\Omega}^3=-\Omega^1 \Omega^2,
\end{equation}
describing the Euler top, and the Darboux--Halphen system met in Eqs. (\ref{halp}). The Lagrange system leads to the Eguchi--Hanson  \cite{Eguchi:1978gw} and Belinsky \emph{et al} solutions \cite{Belinsky:1978ue}, whereas the Darboux--Halphen solutions are the Taub--NUT \cite{Taub-nut} and the Atiyah--Hitchin instantons \cite{Atiyah1}.

The appearance of the $S^3$ Ricci-flow equations  (\ref{halp}), as equations describing the Euclidean-time evolution inside a gravitational instanton foliated with three-spheres is not a coincidence. One can show \cite{Bourliot:2009fr} that it holds for \emph{all} unimodular (class A) Bianchi groups. More generally, one shows that, under the previous assumptions of homogeneous foliation, equations 
(\ref{1stord}) and (\ref{flat}) for the metric (\ref{met}) are geometric-flow equations on $\mathcal{M}_3$, where the flow is driven by a combination of the Ricci tensor and a bilinear of a non-flowing $SU(2)$ Yang--Mills flat gauge connection $a_i$ over the Bianchi group $G$: 
\begin{eqnarray}
\nonumber
\frac{\mathrm{d} \gamma_{ij}}{\mathrm{d}t}=-R_{ij}[\gamma] -\frac{1}{2}  \mathrm{tr} \left( a_i  a_j\right),\\
\nonumber
F\equiv \mathrm{d}a +[a,a]= 0,\quad \frac{\mathrm{d}a}{\mathrm{d}t}= 0.
\end{eqnarray}
The relation between $g_{ij}$ (in (\ref{met})) and $\gamma_{ij}$ (in (\ref{met3})) is as follows:
\begin{equation}
\nonumber
  g_{ij}=\gamma_{ik} \mathcal{K}^{k\ell}\gamma_{\ell j},
\end{equation}
where $ \mathcal{K}^{k\ell}$ is an appropriately chosen invertible tensor. For Bianchi IX, e.g.,  $ \mathcal{K}^{k\ell}$  is the Cartan--Killing metric\footnote{The relationship among gravitational-instanton dynamics and geometric flows, holds in the present homogeneous-space framework as a consequence of a remarkable identity for the curvatures: $
        R[g]= R[\gamma]^2 - \frac{1}{2}R_{ij}[\gamma]\Gamma^{ik}\Gamma^{j\ell} R_{k\ell}[\gamma]
 - \frac{1}{2}R_{ij}[\gamma]\Gamma^{jk}  \mathcal{C}_{k\ell} \Gamma^{\ell m} R_{m n}[\gamma] \mathcal{K}^{ni}$, where $\Gamma^{jk} $ is the inverse of $\gamma_{ij}$, and $\mathcal{C}_{jk}$ the inverse of $\mathcal{K}^{ij}$.}.
 

\section{Concluding remarks and extensions}

Geometric flows appear in a variety of physical problems related to gravity and string theory. In string theory, Ricci or other geometric flows describe the renormalization-group evolution on the worldsheet theory and can, in some regimes where the dilaton is time-dependent,  be identified with genuine time evolution. Besides any practical application, this property raises  the fundamental question of the origin of time in string theory along the (mysterious) line of thought of A. Polyakov. 

In general relativity, Ricci flows (with or without flat $SU(2)$ gauge fields) appear to describe the evolution of the geometry of the homogeneous leaves (spatial sections) inside a self-dual gravitational  instanton. This property emerges in a very constraint framework, where all ingredients (four dimensions, foliation, homogeneity and self-duality) seem to play a crucial  r\^ole. Introducing a cosmological constant is achieved by imposing the self-duality on the Weyl tensor with an immediate consequence at the level of the geometric-flow interpretation: the $SU(2)$ Yang--Mills field on $\mathcal{M}_3$ is no longer flat and the gauge connection flows with time.  The evolution of the leaves is now governed by a  coupled Ricci and Yang--Mills flow.

Geometric flows, i.e. first-order equations, do also describe some gravitational instantons in the non-relativistic theories of gravity  \cite{Horava:2008ih,Horava:2009uw}. The framework provided by the latter is very different from the situations described previously. Indeed, in these theories, the diffeomorphism invariance is \emph{explicitly} broken down to the foliation-preserving transformations, whereas this breaking is \emph{spontaneous} in the example I presented in general relativity.  In the theory of  non-relativistic gravity at hand, we do not impose any self-duality condition on the solution but instead a detailed-balance condition at the level of the potential of the theory, which sets the dynamics by means of an energy--momentum tensor in one dimension less.  With these ingredients, the ground states of the theory, \emph{in any dimension $D+1$}, satisfy first-order flow equations in $D$ dimensions. The precise nature of the flow depends on the $D$-dimensional dynamics  via the detailed-balance condition. For instance, for $D=3$ with a three-dimensional dynamics governed by Einstein--Hilbert action with three-dimensional gravitational coupling $\kappa_W$ and cosmological
constant $\Lambda_W$, plus Chern--Simons action with coupling $w_{\mathrm{CS}}$, the ground states are captured by the following flow equations for the metric (\ref{met}):
\begin{equation}
\frac{\mathrm{d}g_{ij}}{\mathrm{d}t}= -
\frac{\kappa^2}{\kappa_W^2}\left(
R_{ij} - {2\lambda - 1 \over 2(3\lambda -1)} R g_{ij} +\frac{\Lambda_W}{1-3\lambda}  g_{ij}
\right)+
\frac{\kappa^2 }{w_{\mathrm{CS}}} C_{ij}, 
\end{equation}
where $\kappa$ is Newton's constant in four dimensions, and $\lambda$ measures the deviation from the relativistic situation ($\lambda=1$) in the kinetic term. At each flow line, under some regularity conditions (finiteness of the action), corresponds a gravitational instanton (details on the flows  and on their interpretation as gravitational instantons are available in \cite{Bakas:2010fm}).

 \section*{Acknowledgements}
 The material presented in the CORFU 2009 lectures was based, among others, on works made in collaboration with I. Bakas, F. Bourliot, J. Estes, D. L\"ust, D. Orlando and Ph. Spindel. I also benefited from discussions with J.--P. Derendinger, N. Prezas and K. Sfetsos.
 This research was supported by the ERC Advanced Grant  226371,
the ITN programme PITN-GA-2009-237920 , 
the GRC APIC PICS-Gr\`ece  3747 and the IFCPAR CEFIPRA programme 4104-2.

\end{document}